\documentclass[3p,times,twocolumn]{elsarticle}

\usepackage{ecrc}

\volume{00}

\firstpage{1}

\journalname{Nuclear Physics B Proceedings Supplement}

\runauth{S. Ghosh}

\jid{nuphbp}

\jnltitlelogo{Nuclear Physics B Proceedings Supplement}

\usepackage{amssymb}

\usepackage[figuresright]{rotating}

\begin{document}

\begin{frontmatter}



\dochead{}

\title{Measurement of the muon charge asymmetry in inclusive $pp \rightarrow W + X$ production at $\sqrt{s} = 7$ TeV at CMS and an improved determination of light parton distribution functions
}
\author{Saranya Samik Ghosh (for the CMS Collaboration)\fnref{label2}}
\ead{saranya.ghosh@cern.ch}
\fntext[label2]{Speaker}
\address{TIFR, Homi Bhabha Road, Mumbai-5, INDIA}

\dochead{}





\begin{abstract}

Measurements of the muon charge asymmetry in inclusive $pp \rightarrow WX$ production at $\sqrt{s}=7$ TeV are presented. The data sample corresponds to an integrated luminosity of 4.7 $\mathrm{fb^{-1}}$ recorded with the CMS detector at the LHC. With a sample of more than twenty million $W \rightarrow \mu\nu$ events, the statistical precision is greatly improved in comparison to previous measurements. These new results provide additional constraints on the parton distribution functions of the proton in the range of the Bjorken scaling variable $x$ from $10^{-3}$ to $10^{-1}$. These measurements are used together with the cross sections for inclusive deep inelastic ep scattering at HERA in a next-to-leading-order QCD analysis.

\end{abstract}




\end{frontmatter}


\section{Introduction}
\label{sec1Intro}

In the proton-proton collisions at the Large Hadron Collider (LHC), the primary processes for inclusive W-boson production are the annihilation processes: $u\bar{d} \rightarrow W^{+}$ and $d\bar{u} \rightarrow W^{-}$, where a valence quark comes from one proton and a sea quark comes from the other proton. Due to the presence of two valence u quarks as opposed to only one valence d quark in the proton, more $W^{+}$ bosons are crated compared to $W^{-}$ bosons. The measurement of the asymmetry in the production of $W^{+}$ and $W^{-}$ bosons as a function of boson rapidity can provide constraints on the d/u ratio and the sea quark densities in the proton. 
\\
\\
There are large uncertainties on the parton distribution functions (PDFs) for the scale and Bjorken $x$ region explored in the p-p collisions at LHC at $\sqrt{s} = 7$ TeV corresponding to Bjorken $x$ from $10^{-3}$ to $10^{-1}$. In addition, there are differences between different PDF sets. A high precision measurement of the W-boson charge asymmetry can contribute to the improvement of the knowledge of PDFs.
\\
\\
It is difficult to measure the W-boson production asymmetry in terms of the boson rapidity because of the presence of neutrinos in the leptonic W decay. Instead, the lepton charge asymmetry in terms of lepton pseudorapidity is measured. The lepton charge asymmetry is defined as 
\begin{equation}
\mathcal{A}(\eta) = \frac{\frac{d\sigma}{d\eta}{(W^{+} \rightarrow \ell^{+}\nu)} - \frac{d\sigma}{d\eta}{(W^{-} \rightarrow \ell^{-}\bar{\nu})}}{\frac{d\sigma}{d\eta}{(W^{+} \rightarrow \ell^{+}\nu)} + \frac{d\sigma}{d\eta}{(W^{-} \rightarrow \ell^{-}\bar{\nu})}}          
\end{equation}
where $\frac{d\sigma}{d\eta}$ is the differential cross section for W-boson production and subsequent leptonic decay and $\eta=-ln [tan (\theta/2)]$ is the charged lepton pseudorapidity in the laboratory frame, with $\theta$ being the polar angle measured with respect to the beam axis.
\\
\\
The measurement of lepton charge asymmetry in the muon channel, corresponding to the $W \rightarrow \mu\nu$ decay channel of the W-boson, using a data sample corresponds to an integrated luminosity of 4.7 $\mathrm{fb^{-1}}$ recorded with the Compact Muon Solenoid (CMS) detector \cite{bib2CMS} is presented along with a next-to-leading order QCD analysis using the CMS measurement combined with the cross sections for deep inelastic $e^{+-}p$ scattering at HERA \cite{bib3HERA}.

\section{Measurement of muon charge asymmetry}
\label{sec2Measurement}	  

\begin{figure}[htbp]
\centering
\includegraphics[scale=.34]{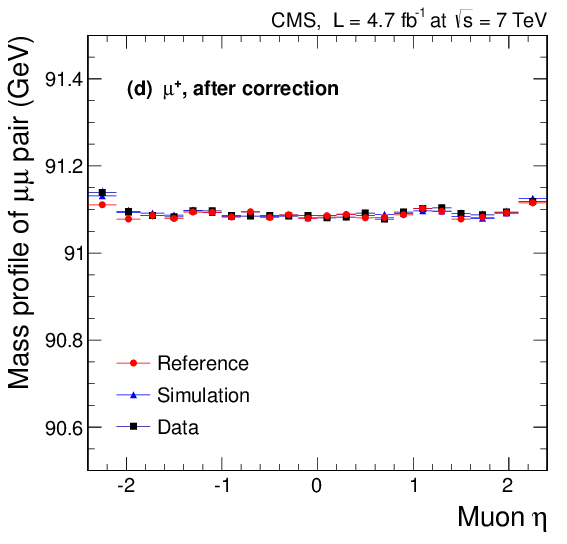}
\caption{
The dimuon mass profile as a function of muon $\eta$ for $\mu+$ after the correction. The reference is obtained using MC generator level information and taking into account the effect of reconstruction on the resolution.
}
\label{fig:fig_muonscalecorn}
\end{figure}

The events selected are required to contain one well reconstructed isolated muon with transverse momentum ($p_{T}$) greater than 25 GeV. An additional measurement of muon charge asymmetry is also made with a higher muon $p_{T}$ threshold of 35 GeV. Events containing a second isolated muon with $p_{T}$ greater than 15 GeV are rejected, in order to reduce the Drell-Yan background. The important backgrounds constitute of Drell-Yan events, QCD multijet events with muons in the final state, $W \rightarrow \tau\nu$ with the $\tau$ decaying into a muon and neutrino and $t\bar{t}$ events. Monte Carlo (MC) simulated samples, that include a full simulation of the CMS detector, are used to help evaluate the background contributions in the data sample.
\\
\\
The selected events are divided into eleven bins based on the absolute pseudorapidity ($|\eta|$) of the muon, corresponding to [0.0-0.2], [0.2-0.4], [0.4-0.6], [0.6-0.8], [0.8-1.0], [1.0-1.2], [1.2-1.4], [1.4-1.6], [1.6-1.85], [1.85-2.1], [2.1-2.4] and the muon charge asymmetry is calculated separately in these eleven bins of muon $|\eta|$.
\\
\\
Corrections are applied in order match the pileup, that is multiple proton-proton interactions in the same event, in MC to that in data (the MC simulation is generated with a different pileup distribution than that we observe in the data) and also to match the boson transverse momentum distribution in MC to that in data. The muons momentum scale is corrected for any bias due to misalignment and mismodelling of the magnetic field. Corrections are based on the charge, pseudorapidity and the azimuthal angle of the muons and have been derived on $Z \rightarrow \mu\mu$ selected sample. The effect of muon momentum scale corrections on positive muons is shown in Figure ~\ref{fig:fig_muonscalecorn}.
 
\begin{figure}[htbp]
\centering
\includegraphics[scale=.34]{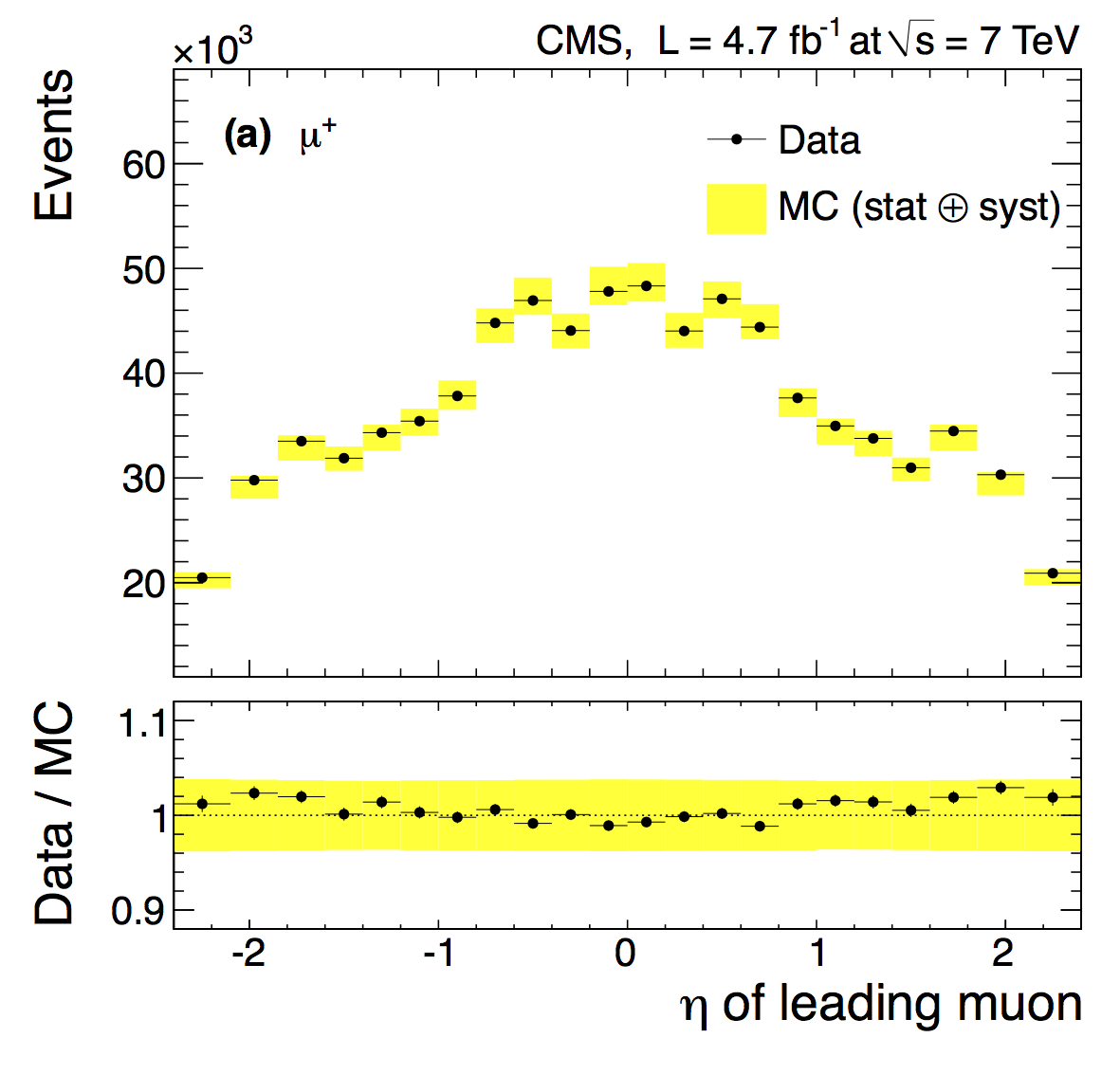}
\caption{
The $\eta$ distribution of the leading $\mu+$ in $Z/\gamma^{*} \rightarrow \mu+ \mu−$ sample. The dimuon invariant mass is within $60 < m_{\mu\mu} < 120$ GeV. The MC simulation is normalized to the data.
}
\label{fig:fig_dynorm}
\end{figure}

To account for the Drell-Yan background, the Drell-Yan MC sample is normalized with theoretical cross-section to luminosity of the data sample. MC to data scale factors are used to account for the difference in efficiency of muon identification and selection between data and MC. Additional k-factors have been calculated to correct normalization. The correction factors are obtained using $Z \rightarrow \mu\mu$ samples in bins of dimuon invariant mass. The Drell-Yan normalisation is shown in Figure ~\ref{fig:fig_dynorm}.
\\
 
\begin{figure}[htbp]
\centering
\includegraphics[scale=.3]{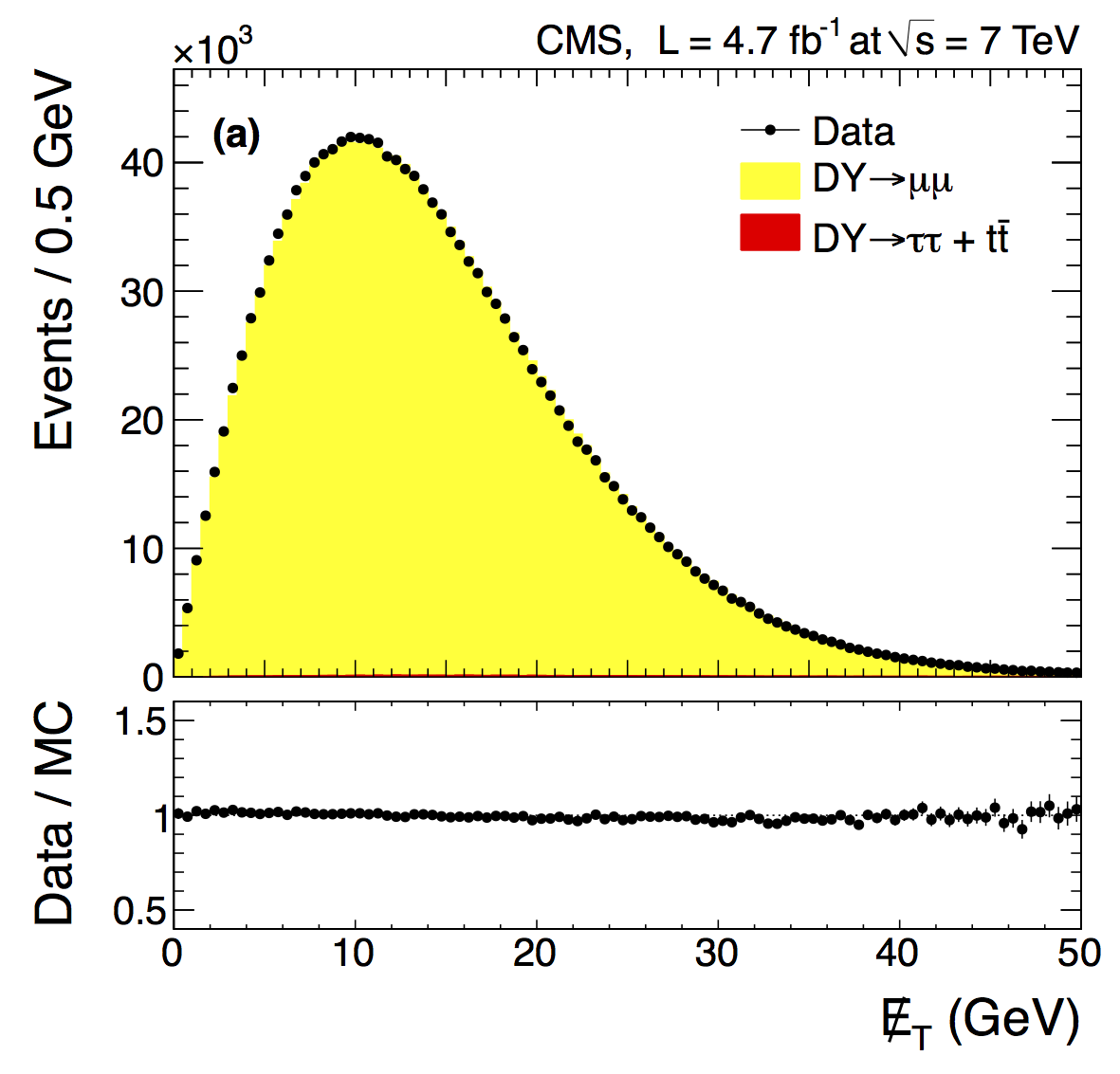}
\caption{
Data to simulation comparison for MET in the Drell–Yan control sample.
}
\label{fig:fig_METdy}
\end{figure}

\begin{figure}[htbp]
\centering
\includegraphics[scale=.34]{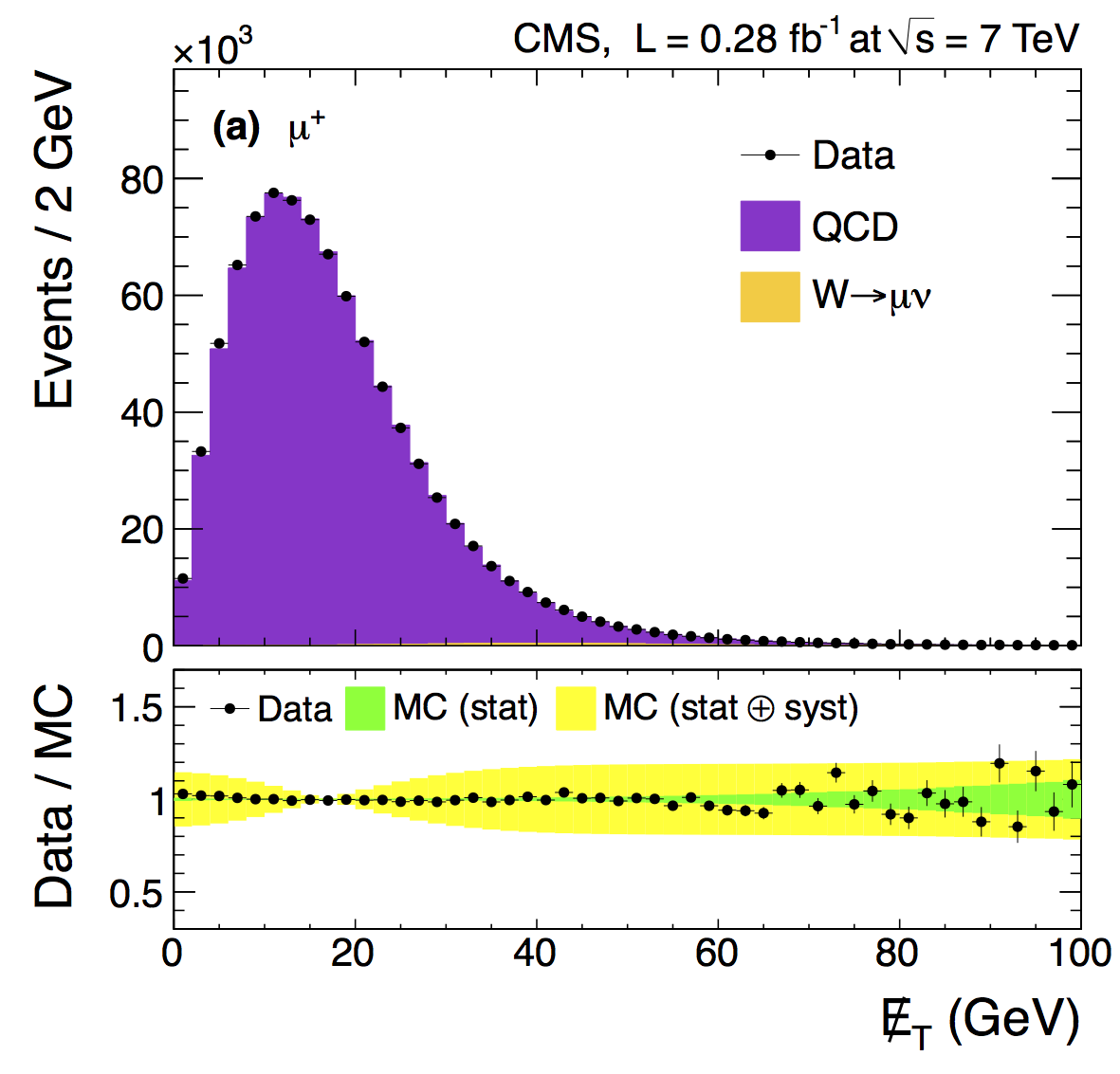}
\caption{
Data to simulation comparison for MET in the QCD control sample for $\mu+$ events. The $W\rightarrow\mu\nu$ contribution (light shaded region) is normalized to the integrated luminosity of the data sample using a MC simulation, and the normalization of the QCD simulation (dark shaded region) is taken as the difference between the data and the estimated $W\rightarrow\mu\nu$ contribution.
}
\label{fig:fig_METqcd}
\end{figure}

The missing transverse energy (MET) distribution is asymmetric in $\phi$ due to detector and reconstruction effects. Corrections are applied to correct for MET $\phi$ asymmetry. Even after correcting for the $\phi$ asymmetry, the distribution of MET in MC is different from that in data. MET is used for signal extraction, and the MET in MC is matched to that in data using recoil corrections. Recoil corrections are corrections to the hadronic recoil, which is the vector sum of transverse momenta of all particle candidates excluding the candidate muon. The corrections are derived on a Drell-Yan selected control sample. The MET distribution in the Drell-Yan selected sample in data and MC after applying the corrections is shown in Figure ~\ref{fig:fig_METdy}. The corrections are verified in a QCD control sample that is obtained by inverting the muon isolation criteria. The MET distribution in the QCD control sample is shown in Figure ~\ref{fig:fig_METqcd}.
\\

\begin{figure}[htbp]
\centering
\includegraphics[scale=.31]{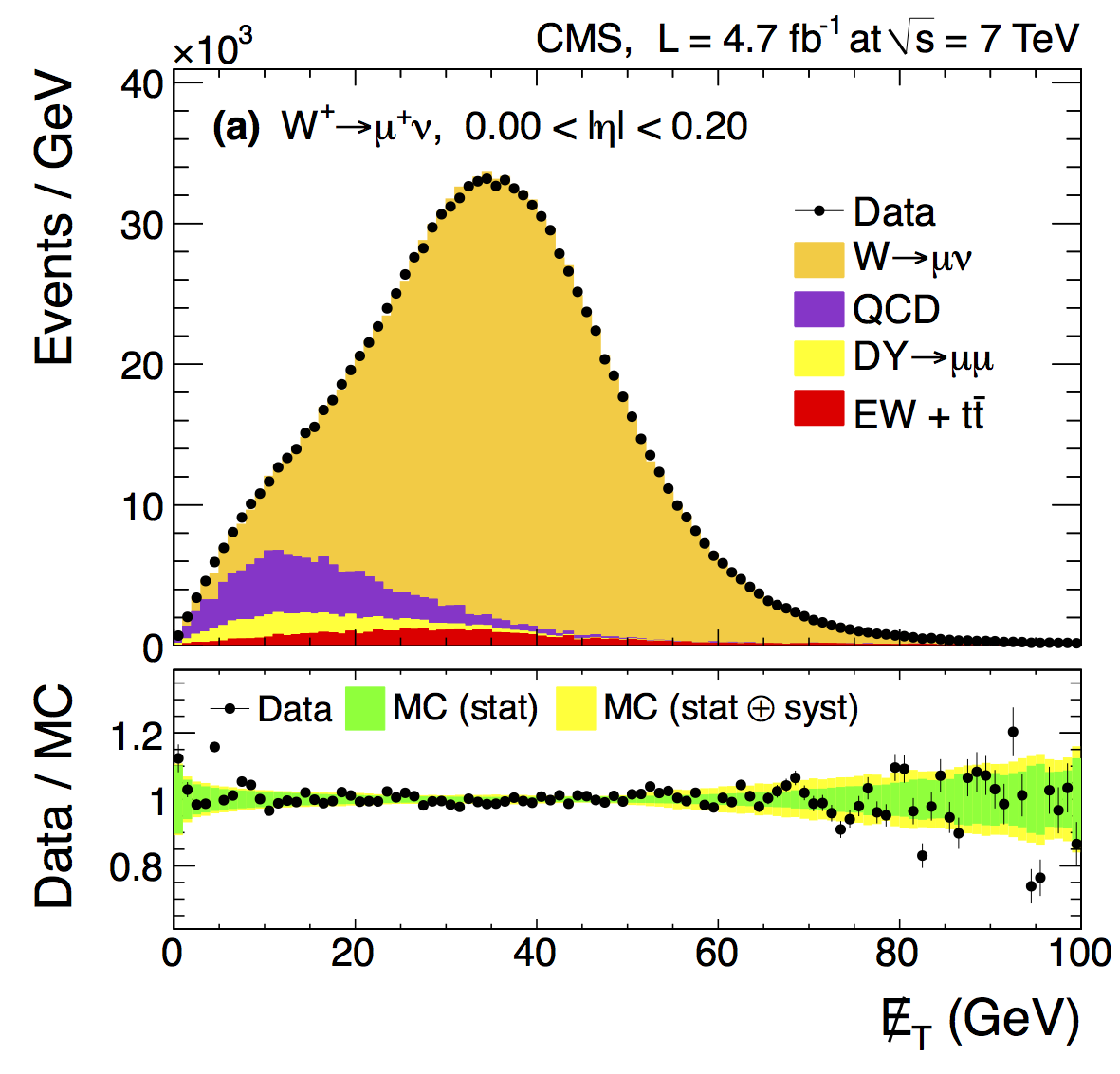}
\includegraphics[scale=.31]{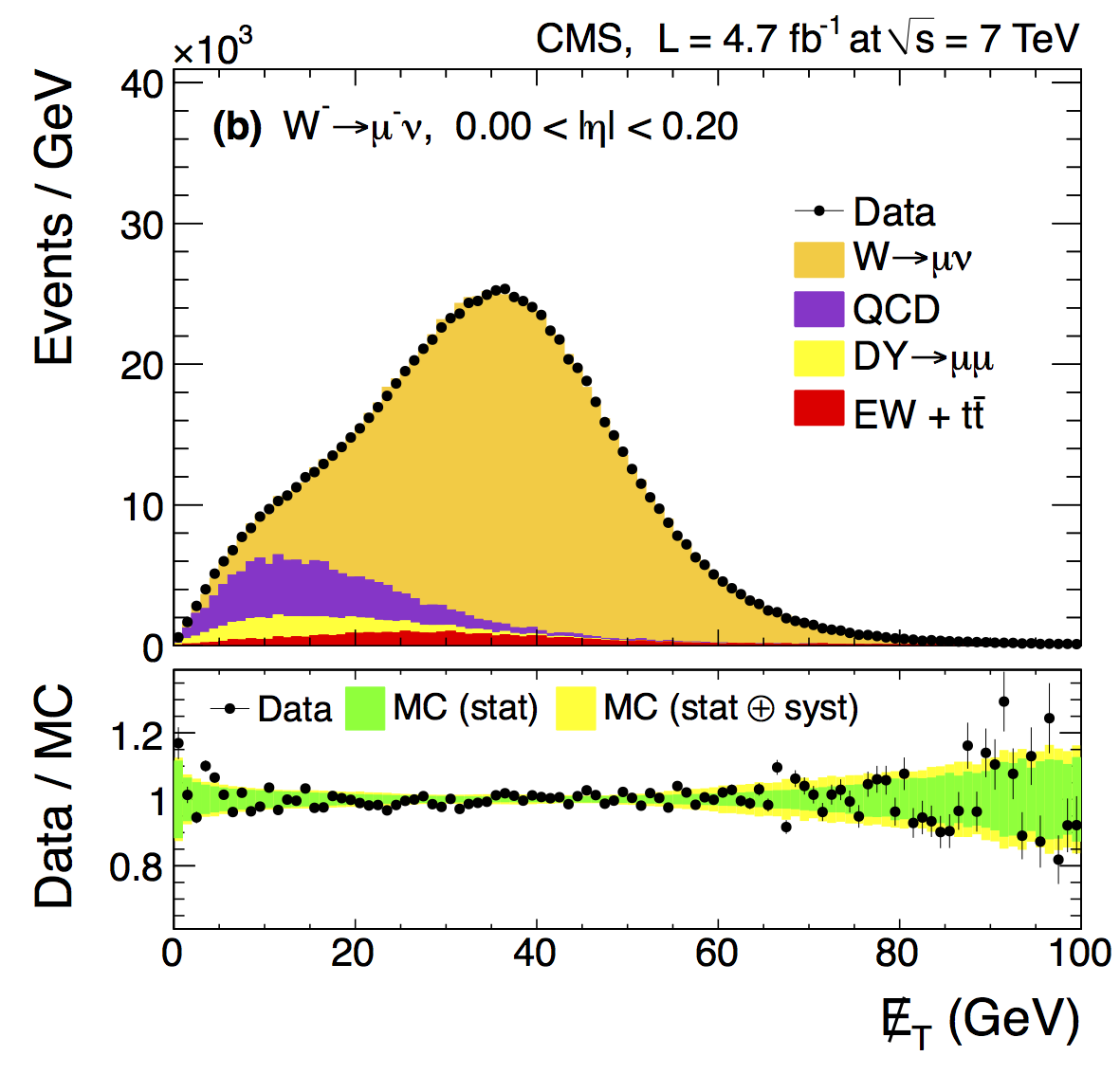}
\caption{
Fits using MET distributions for the muon $p_{T} > 25$ GeV sample for the extraction of the $W\rightarrow\mu\nu$ signal from data for pseudorapidity bin $0.0 \le |\eta| < 0.2$. The fits to $W^{+}\rightarrow\mu^{+}\nu$ (a : top) $W^{-}\rightarrow\mu^{-}\nu$ (b : bottom) are shown separately.
}
\label{fig:fig_fits}
\end{figure}

The process of signal extraction and background estimation is done using the MET distribution in data and MC samples. MET for $W^{+}$ and $W^{-}$ candidate events are taken separately for the different $|\eta|$ bins used and MET templates with corrections from MC fitted to MET distribution in data using binned maximum likelihood fit. Simultaneous but separate fits for $W^{+}$ and $W^{-}$ candidate events is performed in each $|\eta|$ bin with the yields of $W^{+}\rightarrow\mu^{+}\nu$, $W^{-}\rightarrow\mu^{-}\bar{\nu}$ and QCD normalization floated in the fits. A total of 12.9 million $W^{+}$ and 9.1 million $W^{–}$ candidate events give yields of $\sim84\%$ for $W\rightarrow\mu\nu$, $\sim8\%$ QCD, $\sim8\%$ electroweak background consisting of Drell-Yan and $W\rightarrow\tau\nu$ and a small fraction of $t\bar{t}$. An example of the MET fits performed for the muon $p_{T} > 25$ GeV sample for pseudorapidity bin $0.0 \le |\eta| < 0.2$ is shown in Figure ~\ref{fig:fig_fits}.
\\

The asymmetry obtained after the signal extraction process is then corrected for difference in efficiency of selection of positive and negative muons.

\section{Results}
\label{sec3Results}

\begin{figure}[htbp]
\centering
\includegraphics[scale=.38]{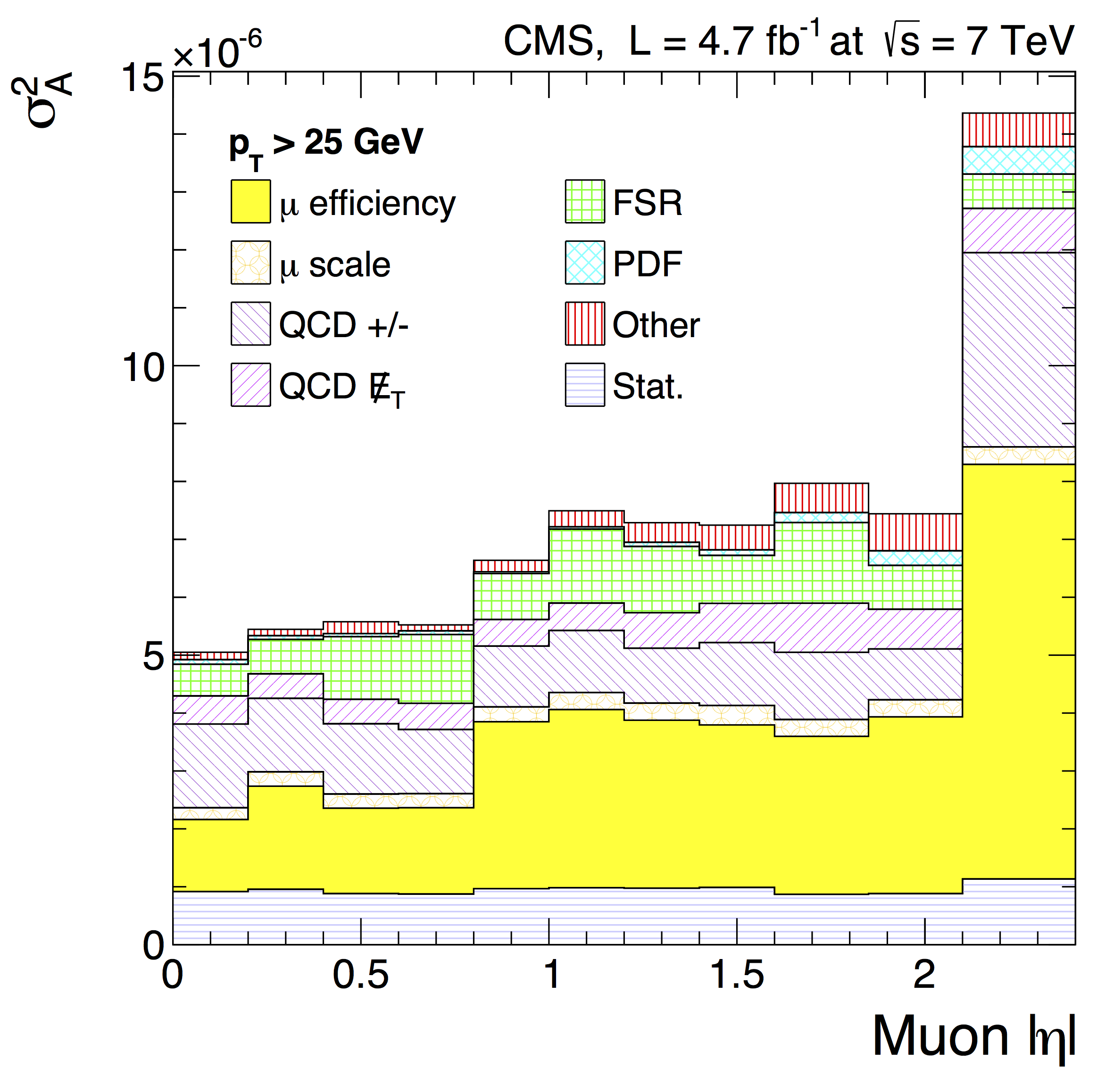}
\caption{
Square of the uncertainties from different sources added in quadrature for uncertainties associated with the $p_{T} > 25$ GeV measurement of muon charge asymmetry for the different $|\eta|$ bins.
}
\label{fig:fig_sysuncert}
\end{figure}

The systematic uncertainties associated with the measured muon charge asymmetry are calculated. The largest source of systematic errors comes from the errors on the measured efficiencies of $\mu^{+}$ and $\mu^{-}$, that are used to correct the asymmetry for difference in efficiency of $\mu^{+}$ and $\mu^{-}$. Another large source of systematic uncertainty comes from the QCD background estimation. The error from the QCD background estimation is taken from two parts, the error coming from the ratio of positive and negative muon events in the QCD background and the uncertainty in the shape of the MET distribution coming from the QCD background. Some of the other major sources of systematic  uncertainty come from uncertainty in the muon momentum scale, effect of final state radiation (FSR) (FSR is not corrected for in the measured asymmetry but systematic uncertainty is assigned to account for its effect) and PDF uncertainties coming from the MC samples used in the measurement. Figure ~\ref{fig:fig_sysuncert} displays the uncertainties coming from different sources for the $p_{T} > 25$ GeV measurement of muon charge asymmetry.
\\

\begin{figure}[htbp]
\centering
\includegraphics[scale=.36]{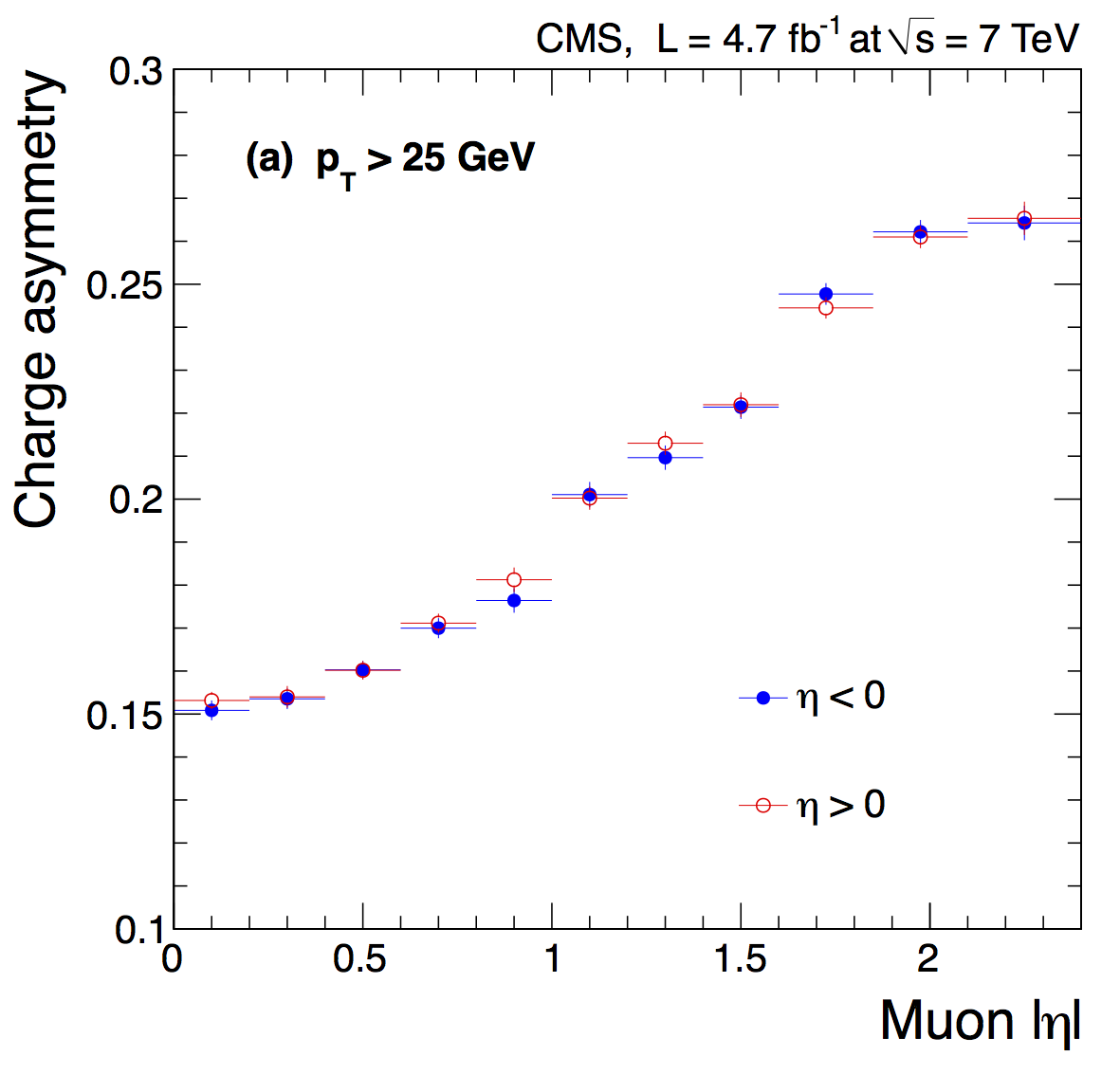}
\caption{
Comparison of the muon charge asymmetry extracted for the positive pseudorapidity ($\eta > 0$) and negative pseudorapidity ($\eta < 0$) regions for muon $p_{T} > 25$ GeV sample. The uncertainties include only the statistical uncertainty
from the signal extraction and uncertainty in the determination of the efficiencies for positive and negative muons.
}
\label{fig:fig_asymetapm}
\end{figure}

As a cross-check, muon charge asymmetry is measured separately for the positive and negative $\eta$ regions by performing an identical measurement in 22 muon $\eta$ bins. The asymmetry obtained from the negative and the positive regions is compared as shown in Figure ~\ref{fig:fig_asymetapm} for the $p_{T} > 25$ GeV case. The charge asymmetries for $\eta > 0$ and $\eta < 0$ regions are in good agreement with each other.
\\

\begin{figure}[htbp]
\centering
\includegraphics[scale=.36]{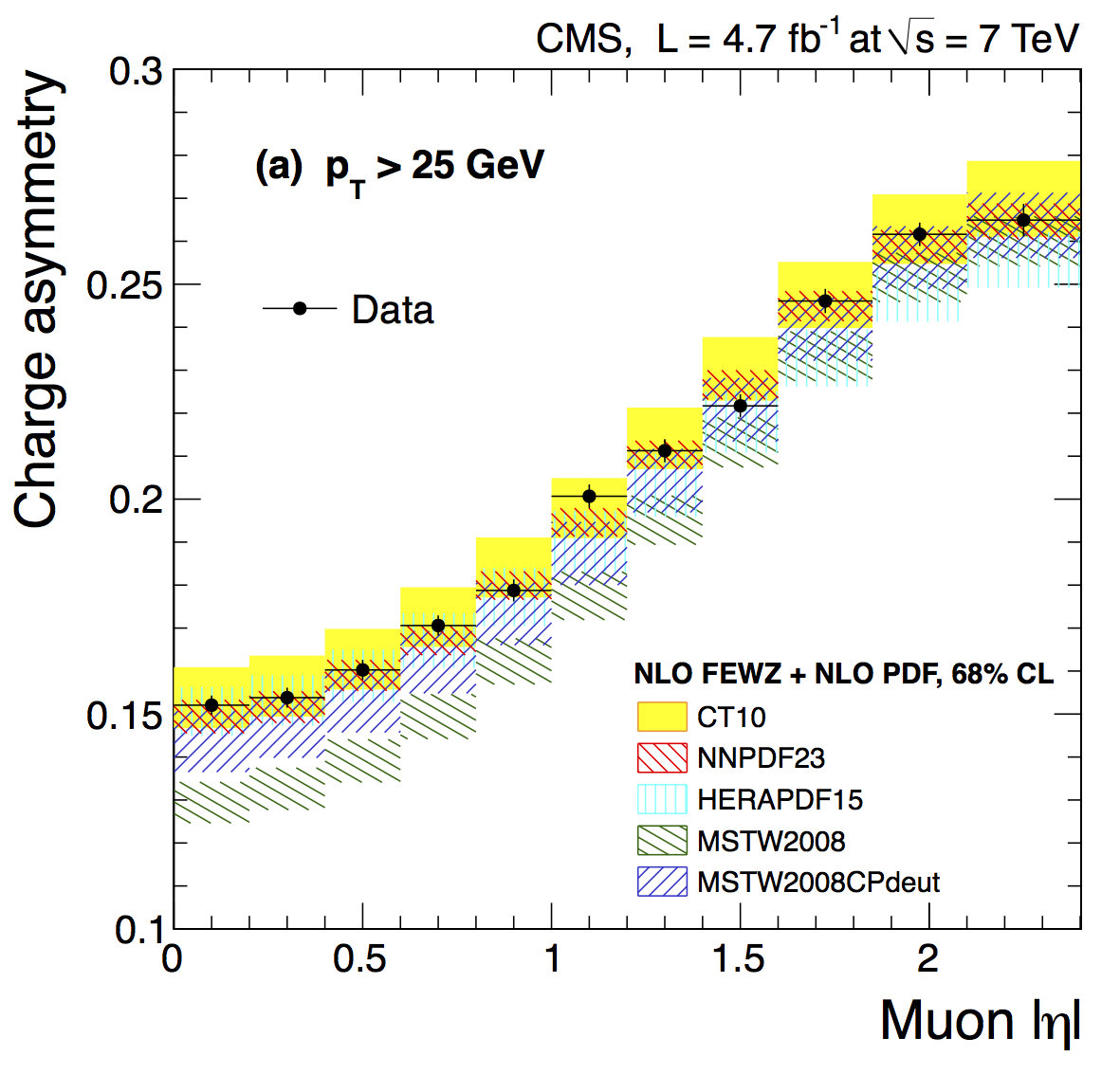}
\caption{
Comparison of the measured muon charge asymmetry for the muon $p_{T} > 25$ GeV sample to the NLO predictions calculated using the FEWZ 3.1 MC tool interfaced with the NLO CT10, NNPDF2.3, HERAPDF1.5, MSTW2008, and MSTW2008CPdeut PDF sets.
}
\label{fig:fig_asym25gev}
\end{figure}

\begin{figure}[htbp]
\centering
\includegraphics[scale=.36]{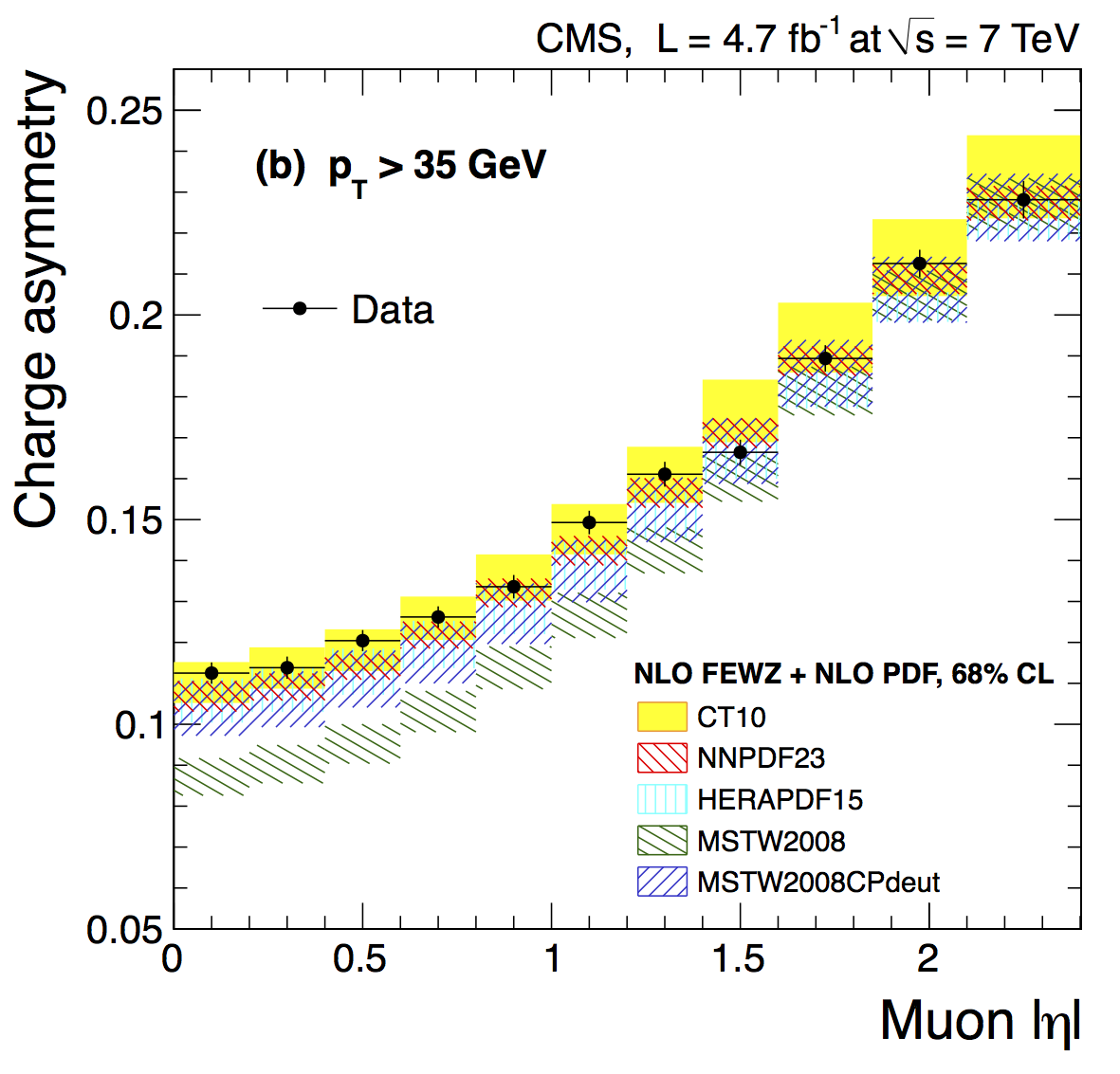}
\caption{
Comparison of the measured muon charge asymmetry for the muon $p_{T} > 35$ GeV sample to the NLO predictions calculated using the FEWZ 3.1 MC tool interfaced with the NLO CT10, NNPDF2.3, HERAPDF1.5, MSTW2008, and MSTW2008CPdeut PDF sets.
}
\label{fig:fig_asym35gev}
\end{figure}

The final measurement of the muon charge asymmetry along with the uncertainties is compared to predictions by five different PDF sets as shown in Figure ~\ref{fig:fig_asym25gev} for the muon $p_{T} > 25$ GeV sample and in Figure ~\ref{fig:fig_asym35gev} for the muon $p_{T} > 35$ GeV sample. The PDF predictions are calculated at NLO using NLO FEWZ 3.1. The measured muon charge asymmetry is in good agreement with predictions from CT10, NNPDF and HERA PDF sets. The agreement with predictions from MSTW PDF set is poor, although it is slightly improved for MSTW2008CPdeut PDF set.
\\

\begin{figure}[htbp]
\centering
\includegraphics[scale=.38]{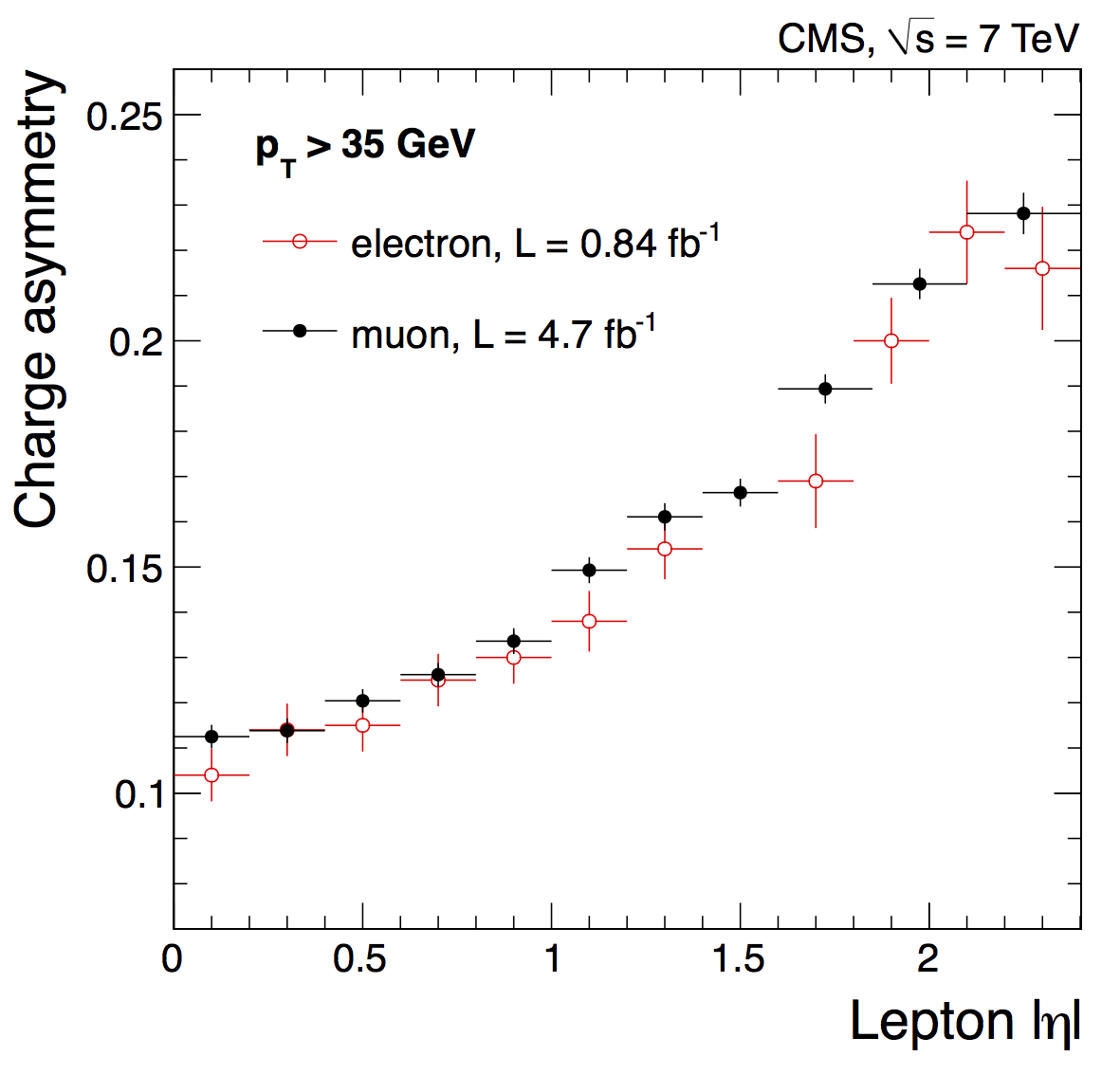}
\caption{
Comparison of measurement of muon charge asymmetry to the previous CMS electron charge asymmetry result. Results are shown for lepton $p_{T} > 35$ GeV.
}
\label{fig:fig_asymcmpelec}
\end{figure}

A comparison of the measured muon charge asymmetry result to the previous CMS electron charge asymmetry measurement \cite{bib4ElecAsym} extracted using 0.84 $\mathrm{fb^{-1}}$ of the 2011 CMS data is shown in Figure ~\ref{fig:fig_asymcmpelec}. The lepton $p_{T}$ threshold on the electron channel measurement is 35 GeV and the comparison is made with the corresponding muon channel result. The muon and the electron charge asymmetry are consistent with each other.
\\

\section{QCD PDF Analysis}
\label{sec4QCDanlysis}

\begin{figure}[htbp]
\centering
\includegraphics[scale=.32]{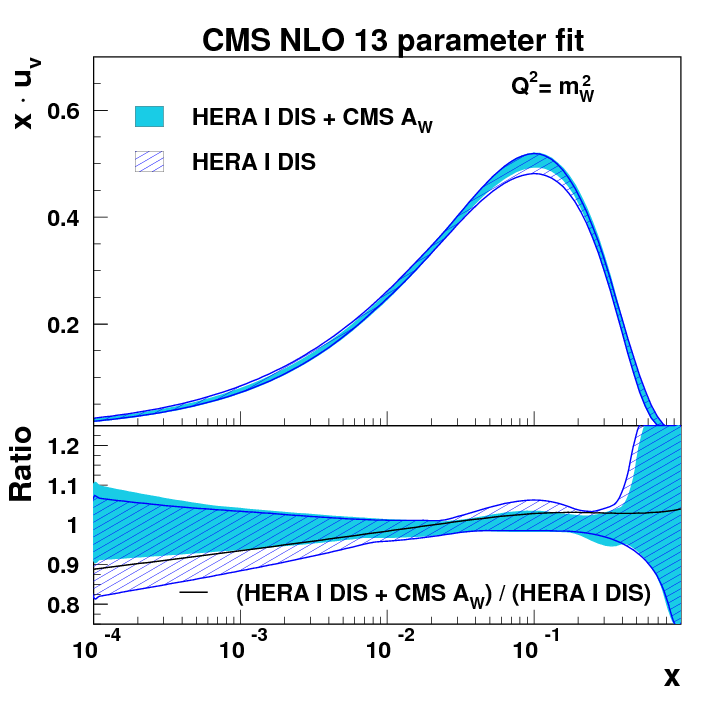}
\includegraphics[scale=.32]{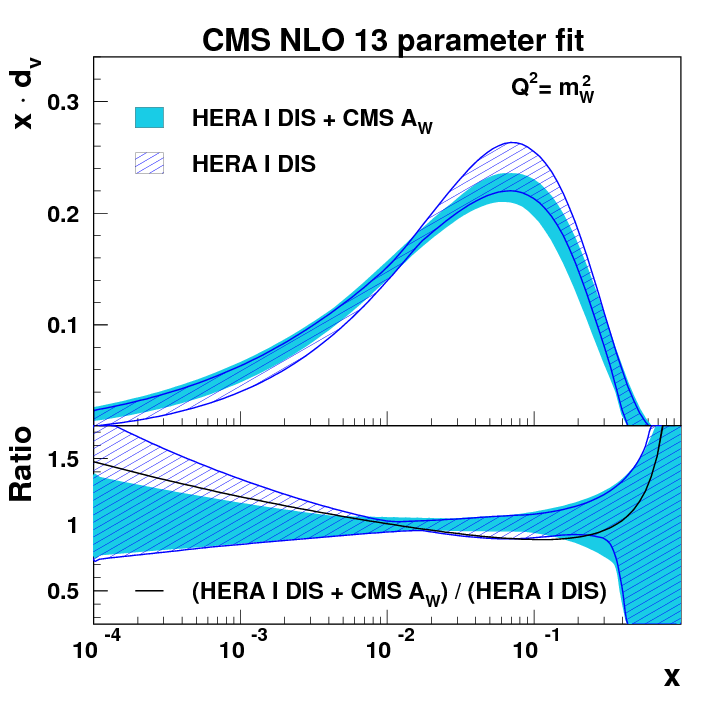}
\caption{
Distributions of $u$ valence (top) and $d$ valence (bottom) quarks as functions of $x$ at the scale $Q^{2} = m_{W}^{2}$ . The PDFs obtained from fit to the HERA data and muon asymmetry measurements (light shaded band), and to HERA only (dark hatched band) are compared. The total PDF uncertainties are shown. In the bottom panels the distributions are normalized to one for a direct comparison of the uncertainties. The change of the PDFs with respect to the HERA-only fit is represented by a solid line.
}
\label{fig:fig_qcdpdfud}
\end{figure}

Combination of CMS muon charge asymmetry measurement and deep inelastic scattering (DIS) data from HERA is used do determine the PDFs of the proton in an NLO perturbative QCD (pQCD) analysis. 
The procedure for the determination of the PDFs follows the approach used in the HERAPDF1.0 QCD fit \cite{bib3HERA}. The PDF uncertainties take into account experimental, model, and parametrization uncertainties according to the general approach of HERAPDF1.0.
\\
\\
A comparison of the PDFs obtained for valence $u$ and $d$ quarks from fits using HERA data and muon asymmetry measurements to that using HERA data only is shown in Figure ~\ref{fig:fig_qcdpdfud}. The muon charge asymmetry measurements, together with HERA DIS cross section data, improve the precision of the PDF of valence quarks over the entire $x$ range.

\section{Conclusion}
\label{sec5Conclusion}

The $W\rightarrow\mu\nu$ lepton charge asymmetry is measured in proton-proton collisions at $\sqrt{s} = 7$ TeV using a data sample corresponding to an integrated luminosity of 4.7 $\mathrm{fb^{-1}}$ collected with the CMS detector
at the LHC. The asymmetry is measured in 11 bins in absolute muon pseudorapidity, $|\eta|$, for two different muon $p_{T}$ thresholds of 25 and
35 GeV. 
\\
\\
The measurement has significantly lower statistical and systematic uncertainties compared to previous CMS measurements. The total uncertainty per bin is $0.2$ to $0.4\%$. The asymmetry measures in data is in good agreement with the theoretical predictions using CT10, NNPDF2.3, and HERAPDF1.5 PDF sets. The agreement with the prediction based on the MSTW2008 PDF set is poor, although the agreement is improved when using the MSTW2008CPdeut PDF set. With smaller uncertainties compared to current PDF uncertainties, this measurement can be used to significantly improve the determination of PDFs in future fits.
\\
\\
An NLO QCD analysis using the muon charge asymmetry along with DIS data from HERA improves the prediction of the valence quark distributions from range $x = 10^{-4}$ to 0.5.




\nocite{*}
\bibliographystyle{elsarticle-num}
\bibliography{martin}

\begin{thebibliography}{00}


\bibitem{bib1pasjme12002} CMS Collaboration, CMS PAS SMP-12-021
\bibitem{bib2CMS} CMS Collaboration,  “The CMS experiment at the CERN LHC”, JINST 3 (2008) S08004
\bibitem{bib3HERA} H1 and ZEUS Collaboration, “Combined measurement and QCD analysis of the inclusive $e^{+-}p$ scattering cross sections at HERA”, JHEP 01 (2010) 109
\bibitem{bib4ElecAsym} CMS Collaboration, “Measurement of the Electron Charge Asymmetry in Inclusive W Production in pp Collisions at $\sqrt{s} = 7$ TeV”, Phys. Rev. Lett. 109 (2012) 111806

\end{thebibliography}



\end{document}